\documentclass[12pt]{article}
\usepackage{epsfig}
\usepackage{amsfonts}
\usepackage{amsmath}
\usepackage{amssymb}
\begin{document}
\title{Multi-channel analog of the effective-range expansion}
\author{
S. A. Rakityansky$^{1)}$
and
N. Elander$^{2)}$\\
$^{1)}${\small\it Dept. of Physics, University of Pretoria, Pretoria 0002, South Africa}\\
$^{2)}${\parbox{12cm}{\small\it Div.of Molecular Physics,
        Dept. of Physics, Stockholm University,\\
        Stockholm, SE-106 91, Sweden}}
}
\maketitle
\begin{abstract}
\noindent
Similarly to the standard effective range expansion that is done near the threshold energy, we obtain a generalized power-series expansion of the multi-channel Jost-matrix that can be done near an arbitrary point on the Riemann surface of the energy within the domain of its analyticity. In order to do this, we analytically factorize its momentum dependencies at all the branching points on the Riemann surface. The remaining single-valued matrix functions of the energy are then expanded in the power-series near an arbitrary point in the domain of the complex energy plane where it is analytic. A systematic and accurate procedure has been developed for calculating the expansion coefficients. This means that near an arbitrary point in the domain of physically interesting complex energies it is possible to obtain a semi-analytic expression for the Jost-matrix (and therefore for the S-matrix) and use it, for example, to locate the spectral points (bound and resonant states) as the S-matrix poles.
\end{abstract}
\vspace{.5cm}
\noindent
PACS number(s): {03.65.Nk, 03.65.Ge, 24.30.Gd}\\[5mm]
Published in:\quad
{\large\bf J.Phys.A44(2011)115303}
\section{Introduction}
Taylor-type power-series expansions are very common in physics. In quantum
scattering theory, the most frequently used expansion of this kind is known as
the effective-range expansion. In the case of a short-range potential, it
represents the cotangent function of the scattering phase shift $\delta_\ell(k)$
in the form
\begin{equation}
\label{effr_L}
   k^{2\ell+1}\cot\delta_{\ell}(k)=\sum_{n=0}^\infty c_{\ell n}k^{2n}\ .
\end{equation}
Here the right-hand side is a sum of terms proportional to even powers of the
collision momentum $k$, and $c_{\ell n}$ are energy independent expansion
coefficients. Originally (see, for example, Ref. \cite{bethe} and Ref.
\cite{goldberger} for a historical review), this expansion was suggested in
nuclear physics for the $S$-wave nucleon-nucleon scattering  in the form
\begin{equation}
\label{effr_0}
   k\cot\delta_0(k)=-\frac{1}{a}+\frac12r_0k^2-Pr_0^3k^4+Qr_0^5k^6+\cdots\ ,
\end{equation}
where the first two parameters on the right-hand side, namely, $a$ and $r_0$,
were called the scattering length and the effective radius. The parameters in
the higher terms of this expansion ($P$, $Q$, etc.) were known as the shape
parameters. This name originated from their dependence on the shape of
a chosen $NN$-potential, while $a$ and $r_0$ were mainly determined  by the
strength of the interaction and the range of the force.\\

Originally, some rather inconvenient integral formulae (see, for example, Ref.
\cite{goldberger}) for calculating the parameters on the right hand side of Eq.
(\ref{effr_0}) were suggested. The expansion, however, was mainly used to
parametrize the experimental data by choosing appropriate values for the
first few coefficients. It is still frequently used for such purposes not only
in nuclear physics but also for parametrizing the low-energy collisions between atoms and molecules (see, for example, Refs.
\cite{hwang, buckman, marinescu, arnecke, idziaszek1, idziaszek2}). The
expansion is also useful for simplified analytical description of the scattering
near the threshold energy. For example, in one of the recent works of this type
the effective-range expansion is used to study the causality constraints on
low-energy universality \cite{Hammer}.\\

A workable and rather accurate method for calculating the low-energy parameters $a$, $r_0$, $P$, etc., was developed about fifty years ago
\cite{kynch, levy, dashen, Babikov1, Babikov}
within the variable-phase approach. Recently the same equations were re-derived in Ref. \cite{Hussein} using the technique of canonical quantization. In this method, the low-energy parameters are the asymptotic values ($r\to\infty$) of the corresponding radial functions $a(r)$, $r_0(r)$, $P(r)$, etc., obeying a system of first-order differential equations. At a fixed $r$, these functions give the corresponding low-energy parameters for the potential which is cut off at this value of the radius.\\

The main drawback of using the differential equations for $a(r)$, $r_0(r)$, $P(r)$, etc., is that such a method is inconvenient in the case of potentials supporting bound states. Indeed, with the radius increasing from zero to infinity, the potential grows from nothing to its complete shape and its bound states appear one by one at certain values of $r$. Each new bound state initially appears with zero binding energy which corresponds to infinite scattering length. Therefore the function $a(r)$ has singularities, which cause serious problems when the system of differential equations is solved numerically. This difficulty is avoided if instead of $a(r)$, $r_0(r)$, $P(r)$, etc., we consider the expansion coefficients of the Jost function for the potential cut off at $r$. These coefficients also obey a system of differential equations but do not have any singularities. The standard low-energy parameters are obtained from them via simple algebra \cite{my2009}.\\

The traditional approach has one significant limitation, namely, the
effective-range expansion is only applicable  near the point $k=0$, i.e. when
the energy is close to the threshold. In Ref. \cite{my2009}, the expansion
(\ref{effr_L}) is generalized by writing it as a series of
powers of $(k-k_0)$ with an arbitrary complex $k_0$.  By doing this, it is possible, for example, to explore the
complex $k$-plane in search for resonances. When $k_0\ne 0$, the generalized effective-range expansion is no longer a low-energy approximation. In other words, instead of
making the expansion around the threshold, one can do it around any point in the complex plane.\\

Yet another limitation of the original expansion (\ref{effr_L}) is that it
is formulated for a single-channel problem.  For multi-channel problems, something
similar was also suggested \cite{Ross1,Ross2,Nath,kermode,sprung,Rosenberg}. In
these papers, the authors tried to keep as much similarity with the
single-channel case as possible. Their adherence to the functions of the type
$\cot\delta$ limited the flexibility and clarity of the suggested equations.\\

In the present paper, we use a different approach in generalizing the
effective-range expansion to multi-channel problems. Instead of considering
the channel phase shifts and their cotangents, we look at the problem from a
more general point of view. As a first step, we analyze the structure of the Jost matrix and explicitly factorize the dependencies of its matrix elements on odd powers of the channel momenta. These factors determine all the branching points of the Riemann surface of the energy, while the remaining factors are single-valued holomorphic functions defined on a simple energy plane. These functions are then expanded in power series of $(E-E_0)$, where $E_0$ is an arbitrary complex energy. As a result, we obtain an expression for the Jost matrix such that each of its elements is a product of a nonanalytic "branching" factor and a power series. In this way, we are not limited to
the threshold points, but expand the Jost matrix practically
anywhere on the Riemann surface. When the Jost matrix is obtained in such a
semi-analytic form (the first several terms of the expansion), the $S$ matrix and all
the observables can be easily calculated. Although it is possible, there is no
need in introducing the generalized scattering length and other parameters. The
Jost matrix expansion coefficients are more simple, clear, and convenient for
this purpose.\\

\section{Multi-channel Schr\"odinger problem}
\label{sect.schreq}
Consider a quantum mechanical two-body problem which, after separation of the
motion of its center of mass, is reduced to an effective problem of one body
whose dynamics is governed by the Hamiltonian
\begin{equation}
\label{total_Hamiltonian}
    H=H_0+{\cal U}+h
\end{equation}
consisting of the free-motion part $H_0$, the interaction operator ${\cal U}$,
and the Hamiltonian $h$ that describes the internal dynamics in the moving
body (for example, the internal states of colliding atoms).
Each internal state of the body corresponds to a different channel of the
scattering process. In general, there are infinite number of the internal
states, i.e. the eigenstates of $h$,
\begin{equation}
\label{infinite_spectrum}
    h|n\rangle=E_n|n\rangle\ ,\qquad n=1,2,3,\dots\ \infty
\end{equation}
We assume that only $N$ internal states are important. In other words, we
approximate the internal Hamiltonian by the $N$ terms
\begin{equation}
\label{finite_spectrum}
    h\approx\sum_{n=1}^N|n\rangle E_n\langle n|\ .
\end{equation}
The total Hamiltonian taken in the representation of the
relative coordinate $\vec{r}$ and sandwiched between $\langle n|$ and
$|n'\rangle$, becomes the $N\times N$ matrix
\begin{equation}
\label{matrix_Hamiltonian}
    H_{nn'}=-\delta_{nn'}\frac{\hbar^2}{2\mu_n}\Delta_{\vec{r}}+
    {\cal U}_{nn'}(\vec{r})+E_n\delta_{nn'}\ ,
\end{equation}
where $\mu_n$ is the reduced mass in the channel $n$.
Therefore the eigenvalue equation $H\Psi=E\Psi$ is reduced to a set of coupled
differential equations for the channel wave functions
\begin{equation}
\label{coupled_eigen}
    \left[\frac{\hbar^2}{2\mu_n}\Delta_{\vec{r}}+(E-E_n)\right]
    \psi_n(E,\vec{r})
    =
    \sum_{n'=1}^N{\cal U}_{nn'}(\vec{r})\psi_{n'}(E,\vec{r})\ .
\end{equation}
The eigenenergies $E_n$ of the internal Hamiltonian $h$ are the thresholds for
the corresponding channels. An eigenstate of the Hamiltonian
(\ref{matrix_Hamiltonian}), corresponding to the eigenvalue $E$, is a
column-matrix
\begin{equation}
\label{column_Psi}
    \Psi(E,\vec{r})=\left(
    \begin{array}{c}
    \psi_1(E,\vec{r})\\
    \psi_2(E,\vec{r})\\
    \vdots\\
    \psi_N(E,\vec{r})
    \end{array}\right)\ ,
\end{equation}
where each row describes the relative motion in the corresponding channel.
\\

To avoid unnecessary complicated notation, we assume that the relative motion in
each channel has a single value $\ell_n$ of the angular momentum. This does not
compromise the generality of our consideration. Indeed, we can always treat the
states with different values of angular momentum (or of any other quantum
number) as different channels, but perhaps with the same threshold energies.\\

Therefore the angular dependence of the channel wave function can be factorized
in the simple way
\begin{equation}
\label{partialwave}
    \psi_n(E,\vec{r})=\frac{u_{n}(E,r)}{r}Y_{\ell_nm_n}(\theta,\varphi)\ .
\end{equation}
This reduces Eq. (\ref{coupled_eigen}) to the system of coupled equations for
the radial parts of the channel wave functions,
\begin{equation}
\label{coupled_radial}
    \left[\partial^2_r+k_n^2-\frac{\ell_n(\ell_n+1)}{r^2}\right]
    u_{n}(E,\vec{r})
    =
    \sum_{n'=1}^NV_{nn'}(r)u_{n'}(E,\vec{r})\ ,
\end{equation}
where
\begin{equation}
\label{potmatr}
    V_{nn'}(r)=\frac{2\mu_n}{\hbar^2}
    \int Y^*_{\ell_nm_n}(\theta,\varphi)
    {\cal U}_{nn'}(\vec{r})Y_{\ell_{n'}m_{n'}}(\theta,\varphi)
    \,d\Omega_{\vec{r}}
\end{equation}
and the channel momenta are defined as
\begin{equation}
\label{chmom}
    k_n=\sqrt{\frac{2\mu_n}{\hbar^2}(E-E_n)}\ .
\end{equation}
In what follows, we assume that the interaction potential is non-singular and of short range, i.e. that all its matrix elements are less singular than
$1/r^2$ at the origin and exponentially vanish at infinity. In principle, the class of acceptable potentials can be much wider. This however would require a complicated analysis of the analytic properties of the multi-channel Jost matrix in order to identify the domain of complex energies where the Jost matrix can be analytically continued to. With the exponentially decaying potentials, we can rely on such an analysis given in Ref. \cite{Motovilov} where the domain of analyticity of the multi-channel $T$-matrix and $S$-matrix is rigorously established. Therefore by narrowing the class of the potentials, we sacrifice the generality for the sake of clarity.\\

The boundary conditions for Eqs. (\ref{coupled_radial}) are derived from the
requirement that any physical solution must be regular at the point $r=0$ and
have special behaviour when $r\to\infty$, which is different for bound,
resonant, and scattering states. In Sect. \ref{sect.transformation}, we
consider these conditions in more detail.

\section{Multi-channel Jost matrix}
\label{sect.JostMatrix}
A system of $N$ linear second-order differential equations of the type
(\ref{coupled_radial}) has $2N$ linearly independent column-solutions and only
half of them are regular at the origin (see, for example, Ref. \cite{brand}).
Combining these regular columns in a square matrix, we obtain the so called
fundamental matrix of the regular solutions,
\begin{equation}
\label{fundmatr}
    \Phi(E,r)=
    \begin{pmatrix}
    \phi_{11}(E,r) & \phi_{12}(E,r) & \cdots & \phi_{1N}(E,r)\\
    \phi_{21}(E,r) & \phi_{22}(E,r) & \cdots & \phi_{2N}(E,r)\\
      \vdots & \vdots & \vdots & \vdots \\
    \phi_{N1}(E,r) & \phi_{N2}(E,r) & \cdots & \phi_{NN}(E,r)\\
    \end{pmatrix}\ .
\end{equation}
Any other regular solution can only be a linear combination of the columns of
this fundamental matrix. In particular, any physical solution of Eqs.
(\ref{coupled_radial}) is a linear combination of the columns of the
corresponding fundamental matrix,
\begin{equation}
\label{lincomb}
     \begin{pmatrix}
     u_1\\  u_2\\  \vdots\\  u_N\\
     \end{pmatrix}
     =
     C_1
     \begin{pmatrix}
     \phi_{11}\\  \phi_{21}\\  \vdots\\  \phi_{N1}\\
     \end{pmatrix}
     +
     C_2
     \begin{pmatrix}
     \phi_{12}\\  \phi_{22}\\  \vdots\\  \phi_{N2}\\
     \end{pmatrix}
     +\dots+
     C_N
     \begin{pmatrix}
     \phi_{1N}\\  \phi_{2N}\\  \vdots\\  \phi_{NN}\\
     \end{pmatrix}
\end{equation}
This guaranties its correct behaviour when
$r\to0$. As far as the asymptotic behaviour ($r\to\infty$) is concerned, its
correct form can be achieved by proper choice of the combination coefficients
$C_n$.\\

Since the potential is of a short range, far away from the origin the right hand
sides of Eqs.(\ref{coupled_radial}) vanish and these equations decouple,
\begin{equation}
\label{decoupled_radial}
    \left[\partial^2_r+k_n^2-\frac{\ell_n(\ell_n+1)}{r^2}\right]
    u_{n}(E,\vec{r})
    \approx 0\ ,\quad\text{when\quad $r\to\infty$}\ .
\end{equation}
These are the Riccati-Bessel equations. As a pair of linearly independent
solutions of each of them, we can take the Riccati-Hankel functions
$h^{(\pm)}_{\ell_n}(k_nr)$.\\

Despite the fact that equations (\ref{decoupled_radial}) are decoupled,
we can still treat them as a system. It therefore has $2N$ linearly independent
column-solutions which can be chosen in many different ways. The most convenient
choice is the following set of $2N$ columns grouped in two square matrices
\begin{equation}
\label{inwaves}
    W^{\rm (in)}=
    \begin{pmatrix}
    h^{(-)}_{\ell_1}(k_1r) & 0 & \cdots & 0\\
    0 & h^{(-)}_{\ell_2}(k_2r) & \cdots & 0\\
    \vdots & \vdots & \vdots & \vdots \\
    0 & 0 & \vdots & h^{(-)}_{\ell_N}(k_Nr)\\
    \end{pmatrix}
\end{equation}
\begin{equation}
\label{outwaves}
    W^{\rm (out)}=
    \begin{pmatrix}
    h^{(+)}_{\ell_1}(k_1r) & 0 & \cdots & 0\\
    0 & h^{(+)}_{\ell_2}(k_2r) & \cdots & 0\\
    \vdots & \vdots & \vdots & \vdots \\
    0 & 0 & \vdots & h^{(+)}_{\ell_N}(k_Nr)\\
    \end{pmatrix}
\end{equation}
that represent the in-coming and out-going spherical waves in all the channels.
These $2N$ columns form a basis in the space of solutions. In other words, any
column-solution of Eq. (\ref{decoupled_radial}) is a linear combination of
these $2N$ columns. In particular, each column of matrix (\ref{fundmatr}) at
large distances becomes such a combination. The combination coefficients have
two subscripts: one to indicate which column of (\ref{fundmatr}) is expanded
and the other is the summation subscript. Similarly to Eqs.
(\ref{inwaves}, \ref{outwaves}), we can group these coefficients in square
matrices (they depend on the choice of the energy, but do not depend on $r$).
Thus we have
\begin{equation}
\label{phi_ass}
    \Phi(E,r)
    \ \mathop{\longrightarrow}\limits_{r\to\infty}
    \ W^{\rm (in)}(E,r)F^{\rm (in)}(E)+
      W^{\rm (out)}(E,r)F^{\rm (out)}(E)\ ,
\end{equation}
where, by analogy with the single-channel case (see, for example, Ref.
\cite{Taylorbook}),
the energy-dependent matrices $F^{\rm (in/out)}(E)$ can be called Jost matrices.
It is not difficult to show that they determine the $S$-matrix
\begin{equation}
\label{S_matrix}
    S(E)=F^{\rm (out)}(E)\left[F^{\rm (in)}(E)\right]^{-1}
\end{equation}
and thus give complete description of the underlying physical system. The
spectral points $E={\cal E}_n$ (bound states and resonances) are those where
the inverse matrix $\left[F^{\rm (in)}(E)\right]^{-1}$ does not exist, i.e.
the points where
\begin{equation}
\label{spectral}
    \det F^{\rm (in)}({\cal E}_n)=0\ .
\end{equation}
\section{Transformation of the Schr\"odinger equation}
\label{sect.transformation}
At large distances, the fundamental regular matrix $\Phi(E,r)$ is a linear
combination (\ref{phi_ass}) with the $r$-independent coefficient matrices
$F^{\rm (in/out)}(E)$. We can, however, look for $\Phi(E,r)$ in the same form
at any point $r$, but with the coefficient matrices depending on $r$,
\begin{equation}
\label{ansatz}
    \Phi(E,r)\equiv
      W^{\rm (in)}(E,r){\cal F}^{\rm (in)}(E,r)+
      W^{\rm (out)}(E,r){\cal F}^{\rm (out)}(E,r)\ .
\end{equation}
Now, instead of one unknown function $\Phi(E,r)$, we have two unknown functions,
${\cal F}^{\rm (in)}(E,r)$ and ${\cal F}^{\rm (out)}(E,r)$, which therefore
cannot be independent of each other. In principle, we can arbitrarily impose any
(reasonable) additional condition relating them. The most convenient is to
demand that
\begin{equation}
\label{lagrange}
      W^{\rm (in)}(E,r)\frac{\partial}{\partial r}{\cal F}^{\rm (in)}(E,r)+
      W^{\rm (out)}(E,r)\frac{\partial}{\partial r}{\cal F}^{\rm (out)}(E,r)
      =0 \ ,
\end{equation}
which is standard in the theory of differential equations, and is
known as the Lagrange condition within the variation parameters method
\cite{Mathews}. Considering Eq. (\ref{phi_ass}), we see that the Lagrange
condition is certainly satisfied at large distances. Therefore, by imposing it,
we do not change the asymptotic behaviour of the solution.\\

Starting from the coupled-channel radial Schr\"odinger equation
(\ref{coupled_radial}), it is not difficult to obtain the corresponding
equations for the new unknown matrices ${\cal F}^{\rm (in/out)}(E,r)$. To this
end, the ansatz (\ref{ansatz}) is simply substituted into Eq.
(\ref{coupled_radial}). After substitution, the equation is transformed and
simplified, using the following: firstly, the fact that matrices $W^{\rm
(in/out)}(E,r)$
solve Eq. (\ref{decoupled_radial}) (i.e. Eq. (\ref{coupled_radial}) without the
right-hand side); secondly, the Lagrange condition (\ref{lagrange});
thirdly, introducing
the
diagonal matrix of the channel momenta
\begin{equation}
\label{kn_matrix}
     K=\begin{pmatrix}
     k_1 & 0 & \cdots & 0\\
     0 & k_2 & \cdots & 0\\
     \vdots & \vdots & \vdots & \vdots\\
     0 & 0 & \cdots & k_N\\
     \end{pmatrix} \ ;
\end{equation}
and finally using known Wronskian of the Riccati-Hankel functions,
\begin{equation}
\label{wronskian}
     W^{\rm (in)}\left[\partial_rW^{\rm (out)}\right]-
     \left[\partial_rW^{\rm (in)}\right]W^{\rm (out)}=2iK\ .
\end{equation}
The derivation can be found in Ref. \cite{two_channel}.\\

As a result, we obtain the following system of first-order differential
matrix-equations
\begin{eqnarray}
\label{fpeq}
    \partial_r{\cal F}^{\rm (in)} &=&
    -\frac{1}{2i}K^{-1}W^{\rm (out)}V\left[
      W^{\rm (in)}{\cal F}^{\rm (in)}+
      W^{\rm (out)}{\cal F}^{\rm (out)}\right]\ ,\\[3mm]
\label{fmeq}
    \partial_r{\cal F}^{\rm (out)} &=&
    \phantom{+}\frac{1}{2i}K^{-1}W^{\rm (in)}V\left[
      W^{\rm (in)}{\cal F}^{\rm (in)}+
      W^{\rm (out)}{\cal F}^{\rm (out)}\right]\ ,
\end{eqnarray}
which are equivalent to the initial second-order Schr\"odinger equation
(\ref{coupled_radial}). Since these equations are of the first order, the
boundary conditions for them can only be imposed at a single point. A natural
way of doing this is to demand that matrix (\ref{ansatz}) is regular at the
origin. Then
\begin{equation}
\label{fin_equals_fout}
     {\cal F}^{\rm (in)}(E,0)= {\cal F}^{\rm (out)}(E,0)
\end{equation}
because both $h_\ell^{(+)}(z)$ and $h_\ell^{(-)}(z)$ are singular at $z=0$, but
their singularities exactly cancel each other in the combination
\begin{equation}
\label{hp_plus_hm}
     h_\ell^{(+)}(z)+h_\ell^{(-)}(z)\equiv 2j_\ell(z)\ .
\end{equation}
The choice of the common value for the functions in Eq. (\ref{fin_equals_fout})
determines the overall normalization of the fundamental matrix (\ref{ansatz}),
because Eqs. (\ref{fpeq},\ref{fmeq}) are linear and homogeneous. In the
single-channel case, the regular solution is usually normalized in such a way
that it coincides with the Riccati-Bessel function when
$r\to0$ \cite{Taylorbook}. From Eq. (\ref{hp_plus_hm}), it is clear that if we
demand that our approach gives the traditional single-channel solution when $N=1$,
we should use the following boundary conditions for Eqs. (\ref{fpeq},\ref{fmeq})
\begin{equation}
\label{fin_fout_bc}
     {\cal F}^{\rm (in)}(E,0)= {\cal F}^{\rm (out)}(E,0)=\frac12I\ ,
\end{equation}
where $I$ is the diagonal unit matrix.\\

Thanks to the fact that  the in-coming and out-going waves in the fundamental
matrix (\ref{ansatz}) are factorized, it is easy to construct the physical
solutions with proper asymptotics. For example, the bound states at large
distances can have only the out-going waves. This is achieved by finding such
combination coefficients in Eq. (\ref{lincomb}), i.e. in
\begin{equation}
\label{spectral_lincomb}
    u_n=\sum_{n'}\Phi_{nn'}C_{n'}
    =
    \sum_{n'n''}
      \left[W^{\rm (in)}_{nn''}{\cal F}^{\rm (in)}_{n''n'}+
      W^{\rm (out)}_{nn''}{\cal F}^{\rm (out)}_{n''n'}\right]C_{n'}
\end{equation}
that all the columns involving the in-coming waves, disappear when
$r\to\infty$. In other words, for the physical wave function of a bound state,
we have
\begin{equation}
\label{bs_ass}
      \sum_{n'}
      {\cal F}^{\rm (in)}_{nn'}(E,r)C_{n'}
      \ \mathop{\longrightarrow}\limits_{r\to\infty}
      \   \sum_{n'}
      {F}^{\rm (in)}_{nn'}(E)C_{n'} =0\ .
\end{equation}
This is a system of linear homogeneous equations for the unknown
combination coefficients $C_n$. It has a non-trivial solution if and only if the
corresponding determinant is zero. This results in Eq. (\ref{spectral}) which
can have solutions at discrete values of the energy. Actually, the resonant
states are found in the same way, i.e. by solving the same Eq. (\ref{spectral}),
but at complex values of ${\cal E}_n$. This approach therefore offers a unified
way of finding both the bound and resonant states as well as the $S$-matrix
(\ref{S_matrix}) for the scattering states. The corresponding physical wave
functions are obtained with correct asymptotic behaviour (analytically
factorized Riccati-Hankel functions).

\section{Riemann surface}
\label{sect.Riemann}
For a fixed (generally speaking, complex) value of the energy $E$, each of the
$N$ channel momenta (\ref{chmom}) can have two different values
\begin{equation}
\label{chmom_pm}
    k_n=\pm\sqrt{\frac{2\mu_n}{\hbar^2}(E-E_n)}\ ,
    \qquad n=1,2,\dots,N\ ,
\end{equation}
depending on the choice of the sign in front of the square root. All
these momenta are involved as papameters in Eqs. (\ref{fpeq},
\ref{fmeq}). This means that the Jost matrices ${F}^{\rm (in/out)}(E)$ are not
single-valued functions of $E$. At each point $E$, they have $2^N$
different values, for all possible combinations of the signs of $N$ channel
momenta.\\

In complex analysis, the multi-valued functions are treated as single-valued,
but defined on a multi-layered complex surface which is called Riemann surface.
In our case, each layer (sheet) of this surface corresponds to a different
combination of the signs of $N$ channel momenta, and thus the Riemann surface
of the energy consists of $2^N$ sheets.\\

When we move around a threshold point, we go from one layer to another. Indeed,
a point on a circle centered at the threshold $E_n$, can be parametrized as
$E=E_n+\rho\exp(i\varphi)$, where $\rho$ is the distance from $E_n$ and
$\varphi$ is the polar angle. The corresponding channel momentum
\begin{equation}
\label{chmom_polar}
    k_n=\sqrt{\frac{2\mu_n\rho}{\hbar^2}}e^{i\varphi/2}
\end{equation}
changes its sign after one full circle  and comes back to its
initial value after two full circles ($\varphi=4\pi$). This means that the
sheets of the Riemann surface are connected to each other and thus form a
united multi-layer manifold. The threshold points $E_n$ ($n=1,2,\dots,N$) are
the branching points on this manifold. By moving around these points, we can
continuously reach any of the $2^N$ sheets.\\

In principle, we can construct the Riemann surface rather arbitrarily by making
cuts and appropriate connections of the layers. In quantum theory, it is
standard that each layer is cut along the real energy axis. The cut starts at
the branching point and goes to infinity in the positive direction. The edges
of the cuts of different layers are interconnected in such a way that the
corresponding channel momenta appropriately change their signs.\\

\begin{figure}[ht!]
\centerline{\epsfig{file=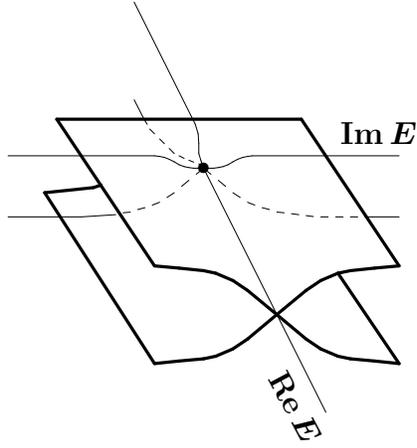}}
\caption{\sf
Riemann surface of the energy for a single-channel problem.
}
\label{fig.Riemann_single}
\end{figure}

\begin{figure}[ht!]
\centerline{\epsfig{file=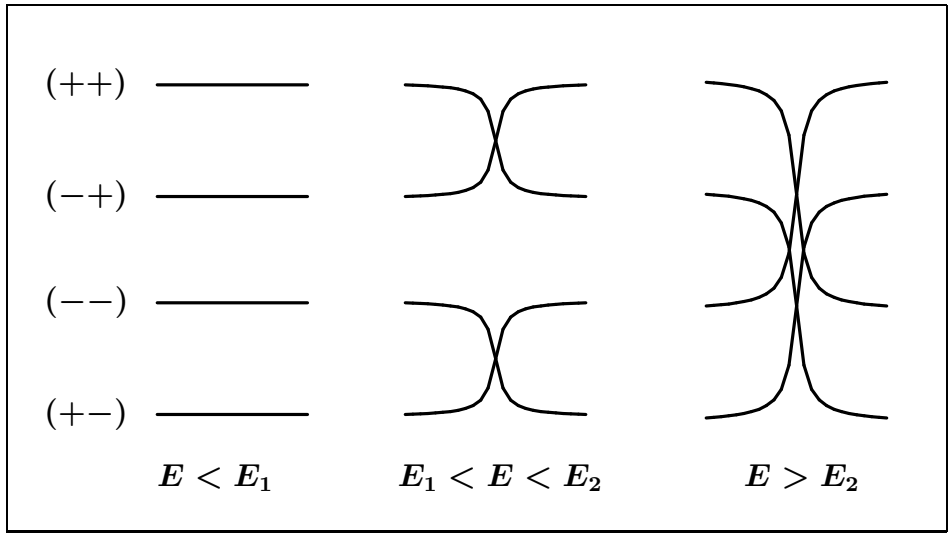}}
\caption{\sf
Schematically shown interconnections of the layers of the Riemann surface for a
two-channel problem at three different energy intervals. The layers
correspond to different combinations of the signs (indicated in brackets) of
${\rm Im}\,k_1$ and ${\rm Im}\,k_2$.
}
\label{fig.Riemann_double}
\end{figure}

The simplest two-layer Riemann surface for the single-channel case is easy to
visualize (see, Fig. \ref{fig.Riemann_single}). The two-channel problem with
two branching points and four interconnected layers is much more involved.
These connections for the three intervals, $E<E_1$, $E_1<E<E_2$, and $E>E_2$ are
schematically shown in Fig. \ref{fig.Riemann_double} \cite{frazer}. When $N>2$,
the surface is so complicated that it is not worthwhile even trying to visualize
it.\\

In the present paper, we construct the Jost matrices in such a way that in their
matrix elements the dependences on odd powers of all channel momenta are
factorized analytically (see Sec. \ref{sec.factor}). The remaining matrices
depend only on even powers of all the momenta $k_n$ and thus are single-valued
functions of variable $E$. This saves us the trouble of dealing with the
complicated multi-layered manifold. Moreover, using the analytically factorized
dependence on $k_n$, we can establish some of the symmetry properties of Jost
matrices, i.e. we can relate their values at some points belonging to different
layers of the Riemann surface (see Sec. \ref{sec.symmetry}).

\section{Complex rotation}
\label{sec.rotation}
When the potential is cut off at certain radius $R$, the right-hand sides
of Eqs. (\ref{fpeq},\ref{fmeq}) vanish for $r>R$ and the derivatives
$\partial_r{\cal F}^{\rm (in/out)}$ become zero, i.e. these functions do not
change beyond this point. Therefore, in the spirit of the variable phase
approach, the functions ${\cal F}^{\rm (in/out)}(E,r)$ are the Jost matrices for
the potential which is cut off at the point $r$. In general, when the potential
asymptotically vanishes at large distances, we
have
\begin{equation}
\label{Flimit}
      {\cal F}^{\rm (in/out)}(E,r)
      \ \mathop{\longrightarrow}\limits_{r\to\infty}
      \  {F}^{\rm (in/out)}(E)\ .
\end{equation}
Therefore, the Jost matrices can be calculated by numerical integration of the
differential equations (\ref{fpeq},\ref{fmeq}) from $r=0$ up to a sufficiently
large radius $R$ where the limit (\ref{Flimit}) is reached within a required
accuracy.\\

\begin{figure}[ht!]
\centerline{\epsfig{file=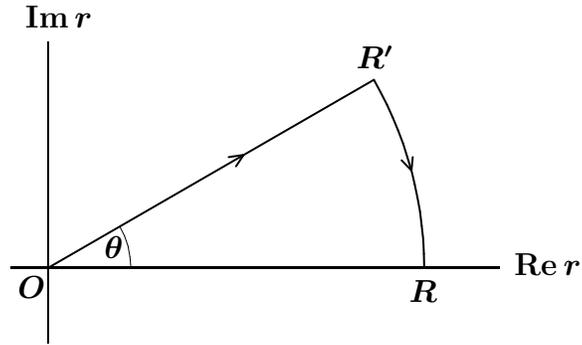}}
\caption{\sf
A deformed path for integrating the differential equations.
}
\label{fig.pathray}
\end{figure}

This works perfectly for real values of the energy $E$. However, when we
consider complex energies (for example, in search for resonances), a technical
difficulty arises. This difficulty is caused by the asymptotic behaviour of the
Riccati-Hankel functions \cite{abramowitz},
\begin{equation}
\label{Hankel_ass}
   h_\ell^{(\pm)}(kr)
   \ \mathop{\longrightarrow}\limits_{|kr|\to\infty}
   \ \mp i\exp\left(\pm ikr\mp i\frac{\ell\pi}{2}\right)\ .
\end{equation}
As is seen, when $k$ is complex, either $h_\ell^{(+)}(kr)$ or $h_\ell^{(-)}(kr)$
exponentially diverges, depending on the sign of ${\rm Im\,}k$.
As a result, either the first or the second of the equations
(\ref{fpeq},\ref{fmeq}) does not give a numerically convergent solution. This
difficulty is circumvented by using the deformed integration path shown in Fig.
\ref{fig.pathray}. Instead of integrating the differential equations along the
real axis from $r=0$ to $r=R$, we can reach the final point via the
intermediate point $r=R'$ in the complex plane.
Moreover, we can safely ignore the arc $R'R$ since the potential is practically
zero at that distance.\\

Why does this complex rotation help? The answer can be found by looking at Eq.
(\ref{Hankel_ass}). Indeed, the asymptotic behaviour (divergent or convergent)
of the functions $h_\ell^{(\pm)}(kr)$ is determined by the sign of ${\rm
Im\,}(kr)$. If $k=|k|e^{i\varphi}$, we can always find such a rotation angle
$\theta$ in $r=|r|e^{i\theta}$ that the product
$$
      kr=|kr|e^{i(\varphi+\theta)}
$$
has either positive or negative (or even zero) imaginary part. Various
technical details of using complex rotation in calculating the Jost functions
and Jost matrices can be found in Refs.
\cite{my01, my02, my03, my04, my05, my06, my07, my08, my09, my10}.

\section{Factorization}
\label{sec.factor}
Eqs. (\ref{fpeq},\ref{fmeq}) are very convenient for numerical calculation of
the Jost matrices. However, for the purpose of power-series expansion of these
matrices, we need to further transform them. The idea of such a transformation
is based on the relation between the two pairs of linearly independent
solutions of the Riccati-Bessel equation, namely, between the Riccati-Hankel
functions $h_\ell^{(\pm)}$ and the pair of Riccati-Bessel $j_\ell$ and
Riccati-Neumann $y_\ell$ functions,
\begin{equation}
\label{hjy}
    h_\ell^{(\pm)}(z)=j_\ell(z)\pm iy_\ell(z)\ .
\end{equation}
Introducing the diagonal matrices
\begin{equation}
\label{Jmatr}
    J=\frac12\left[W^{\rm(in)}+W^{\rm(out)}\right]=
    \begin{pmatrix}
    j_{\ell_1}(k_1r) & 0 & \cdots & 0\\
    0 & j_{\ell_2}(k_2r) & \cdots & 0\\
    \vdots & \vdots & \vdots & \vdots \\
    0 & 0 & \vdots & j_{\ell_N}(k_Nr)\\
    \end{pmatrix}\ ,
\end{equation}
\begin{equation}
\label{Ymatr}
    Y=\frac{i}{2}\left[W^{\rm(in)}-W^{\rm(out)}\right]=
    \begin{pmatrix}
    y_{\ell_1}(k_1r) & 0 & \cdots & 0\\
    0 & y_{\ell_2}(k_2r) & \cdots & 0\\
    \vdots & \vdots & \vdots & \vdots \\
    0 & 0 & \vdots & y_{\ell_N}(k_Nr)\\
    \end{pmatrix}\ ,
\end{equation}
as well as the new unknown matrices
\begin{eqnarray}
\label{Amatr}
    {\cal A}(E,r)
    &=&
    {\cal F}^{\rm (in)}(E,r)+{\cal F}^{\rm (out)}(E,r)\ ,\\[3mm]
\label{Bmatr}
    {\cal B}(E,r)
    &=&
    i\left[{\cal F}^{\rm (in)}(E,r)-{\cal F}^{\rm (out)}(E,r)\right]\ ,
\end{eqnarray}
we obtain another (equivalent) representation of the fundamental matrix of
regular solutions,
\begin{equation}
\label{PhiAB}
    \Phi(E,r)=J(E,r){\cal A}(E,r)-Y(E,r){\cal B}(E,r)\ .
\end{equation}
Combining Eqs. (\ref{fpeq},\ref{fmeq}), it is easy to obtain an equivalent
system of differential equations for the new unknown matrices,
\begin{eqnarray}
\label{Aeq}
    \partial_r{\cal A} &=&
    -K^{-1}YV\left(J{\cal A}-Y{\cal B}\right)\ ,\\[3mm]
\label{Beq}
    \partial_r{\cal B} &=&
    -K^{-1}JV\left(J{\cal A}-Y{\cal B}\right)\ ,
\end{eqnarray}
with the boundary conditions
\begin{equation}
\label{ABbcond}
    {\cal A}(E,0)=I\ ,\qquad {\cal B}(E,0)=0\ ,
\end{equation}
which immediately follow from (\ref{fin_fout_bc}). Similarly to the limit
(\ref{Flimit}), these matrices also should converge to their asymptotic values
\begin{equation}
\label{ABlimit}
      {\cal A}(E,r)
      \ \mathop{\longrightarrow}\limits_{r\to\infty}
      \  A(E)\ ,\qquad
      {\cal B}(E,r)
      \ \mathop{\longrightarrow}\limits_{r\to\infty}
      \  B(E)\ ,
\end{equation}
from which the Jost matrices can be obtained,
\begin{equation}
\label{JmAB}
     F^{\rm(in)}(E)=\frac12\left[A(E)-iB(E)\right]\ ,\qquad
     F^{\rm(out)}(E)=\frac12\left[A(E)+iB(E)\right]\ .
\end{equation}
Now, we use the fact that the Riccati-Bessel and Riccati-Neumann functions can
be represented by absolutely convergent series,
\begin{equation}
\label{jseries}
    j_\ell(kr)=\left(\frac{kr}{2}\right)^{\ell+1}\sum_{n=0}^\infty
    \frac{(-1)^n\sqrt{\pi}}{\Gamma(\ell+3/2+n)n!}
    \left(\frac{kr}{2}\right)^{2n}=k^{\ell+1}\tilde{j}_\ell(E,r)\ ,
\end{equation}
\begin{equation}
\label{yseries}
    y_\ell(kr)=\left(\frac{2}{kr}\right)^{\ell}\sum_{n=0}^\infty
    \frac{(-1)^{n+\ell+1}}{\Gamma(-\ell+1/2+n)n!}
    \left(\frac{kr}{2}\right)^{2n}=k^{-\ell}\tilde{y}_\ell(E,r)\ ,
\end{equation}
where we factorize the the functions $\tilde{j}_\ell$ and $\tilde{y}_\ell$,
which do not depend on odd powers of $k$ and thus are single-valued functions
of the energy $E$.\\

Let us look for the matrices ${\cal A}$ and ${\cal B}$ in the form
\begin{equation}
\label{ABfact}
    {\cal A}_{ij}=\frac{k_j^{\ell_j+1}}{k_i^{\ell_i+1}}
    \tilde{\cal A}_{ij}\ ,
    \qquad
    {\cal B}_{ij}=k_i^{\ell_i}k_j^{\ell_j+1}
    \tilde{\cal B}_{ij}\ ,
\end{equation}
where certain powers of the channel momenta are factorized in each individual
matrix element. When the representations (\ref{jseries}, \ref{yseries},
\ref{ABfact}) are substituted into Eqs. (\ref{Aeq},\ref{Beq}), all the
channel-momenta factors cancel out, and we remain with the equations for the
tilded functions,
\begin{eqnarray}
\label{tAeq}
    \partial_r\tilde{\cal A} &=&
    -\tilde{Y}V\left(\tilde{J}\tilde{\cal A}-
     \tilde{Y}\tilde{\cal B}\right)\ ,\\[3mm]
\label{tBeq}
    \partial_r\tilde{\cal B} &=&
    -\tilde{J}V\left(\tilde{J}\tilde{\cal A}-
     \tilde{Y}\tilde{\cal B}\right)\ ,
\end{eqnarray}
where the matrices $\tilde{J}$ and $\tilde{Y}$ differ from (\ref{Jmatr}) and
(\ref{Ymatr}) by the diagonal factors,
 \begin{equation}
\label{tJYmatr}
    J=
    \begin{pmatrix}
    k_1^{\ell_1+1} & 0 & \cdots & 0\\
    0 & k_2^{\ell_2+1} & \cdots & 0\\
    \vdots & \vdots & \vdots & \vdots \\
    0 & 0 & \vdots & k_N^{\ell_N+1}\\
    \end{pmatrix}\tilde{J}\ ,
    \qquad
    Y=
    \begin{pmatrix}
    k_1^{-\ell_1} & 0 & \cdots & 0\\
    0 & k_2^{-\ell_2} & \cdots & 0\\
    \vdots & \vdots & \vdots & \vdots \\
    0 & 0 & \vdots & k_N^{-\ell_N}\\
    \end{pmatrix}\tilde{Y}\ .
\end{equation}
The main advantage of Eqs. (\ref{tAeq},\ref{tBeq}) is that they do not involve
any coefficients or functions depending on odd powers of the channel momenta.
This means that their solutions, i.e. the matrices $\tilde{\cal A}(E,r)$ and
$\tilde{\cal B}(E,r)$, are single-valued functions of the energy. The
multi-valuedness of the initial matrices ${\cal A}(E,r)$ and ${\cal B}(E,r)$
as well as the fact that they are defined on a complicated Riemann surface, are
determined by the momentum-factors separated in Eqs. (\ref{ABfact}).

\section{Symmetry of the Jost matrices}
\label{sec.symmetry}
As an example of usefulness of the analytic structure of the Jost matrices,
established in the previous section, let us consider the relation between
$F^{\rm (in)}$ and $F^{\rm (out)}$. If $\tilde{A}(E)$ and $\tilde{B}(E)$ are
the asymptotic values of $\tilde{\cal A}(E,r)$ and $\tilde{\cal B}(E,r)$,
respectively, then according to Eqs. (\ref{JmAB}, \ref{ABfact}), we have the
following semi-analytic expressions for the Jost matrices
\begin{eqnarray}
\label{fin_semi}
    F^{\rm (in)}_{mn} &=&
                 \frac{k_n^{\ell_n+1}}{2k_m^{\ell_m+1}}\tilde{A}_{mn}-
                 \frac{ik_m^{\ell_m}k_n^{\ell_n+1}}{2}\tilde{B}_{mn}\ ,\\[3mm]
\label{fout_semi}
    F^{\rm (out)}_{mn} &=&
                 \frac{k_n^{\ell_n+1}}{2k_m^{\ell_m+1}}\tilde{A}_{mn}+
                 \frac{ik_m^{\ell_m}k_n^{\ell_n+1}}{2}\tilde{B}_{mn}\ .
\end{eqnarray}
If we change the signs of all the channel momenta to the opposite, the
matrices  $\tilde{A}(E)$ and $\tilde{B}(E)$ remain unchanged while the
factorized momenta generate a common factor
$(-1)^{\ell_m+\ell_n}=(-1)^{\ell_m-\ell_n}$ and the sign
between the two terms in Eqs. (\ref{fin_semi}, \ref{fout_semi}) is also changed
to the opposite. In other words,
\begin{equation}
\label{FinFoutSym}
      F^{\rm (in)}_{mn}(-k_1,-k_2,\dots,-k_N)=
      (-1)^{\ell_m+\ell_n}F^{\rm (out)}_{mn}(k_1,k_2,\dots,k_N)\ ,
\end{equation}
where for the sake of clarity, we replaced the single independent
variable $E$ with the set of channel momenta. This means that the two Jost
matrices $F^{\rm (in)}$ and $F^{\rm (out)}$ are not completely independent.
Their values at different points on the Riemann surface are the same. The $S$
matrix can be re-written as
\begin{equation}
\label{SmatrSym}
     S_{mn}=(-1)^{\ell_m+\ell_n}
     F^{\rm (in)}_{mn}(-k_1,-k_2,\dots,-k_N)
     \left[
     F^{\rm (in)}_{mn}(k_1,k_2,\dots,k_N)\right]^{-1}\ ,
\end{equation}
which is well known in the case of a single-channel problem. The symmetry
property (\ref{FinFoutSym}) is the simplest one (but perhaps the most
important) and can be established in a different way (see, for example, Ref.
\cite{Taylorbook}). Its simple proof is given here to merely demonstrate how
the factorized semi-analytic expressions (\ref{fin_semi}, \ref{fout_semi}) can
be used. Establishing the other symmetry properties is beyond the scope of the
present paper.

\section{Power-series expansion}
\label{sec.expansion}
The coefficients of differential equations (\ref{tAeq},\ref{tBeq}), i.e. the products of $\tilde{J}(E,r)$, $\tilde{Y}(E,r)$, and the potential matrix $V$, depend on the parameter $E$, but do not involve any functions depending on channel momenta. The boundary conditions for them,
\begin{equation}
\label{tABbcond}
    \tilde{\cal A}(E,0)=I\ ,\qquad \tilde{\cal B}(E,0)=0\ ,
\end{equation}
are also momentum independent. This means that the matrices
$\tilde{\cal A}(E,r)$ and
$\tilde{\cal B}(E,r)$ are single-valued functions of the parameter $E$,
i.e. they are defined on a single sheet of the complex $E$-plane.\\

Since the matrices $\tilde{J}(E,r)$ and $\tilde{Y}(E,r)$ are
holomorphic, the coefficients of differential equations (\ref{tAeq},\ref{tBeq}) are also holomorphic for any finite radius $r$. Moreover, the boundary conditions (\ref{tABbcond}) do not depend on $E$. According to the Poincar\'e
theorem \cite{poincare}, this means that $\tilde{\cal A}(E,r)$ and
$\tilde{\cal B}(E,r)$ are holomorphic functions of the parameter $E$ for any finite value of $r$. Therefore they can be expanded in the power series around an arbitrary point $E_0$ on the complex plane of the energy,
\begin{eqnarray}
\label{PSEA}
     \tilde{\cal A}(E,r) &=& \sum_{n=0}^\infty
     (E-E_0)^n\alpha_n(E_0,r)\ ,\\[3mm]
\label{PSEB}
     \tilde{\cal B}(E,r) &=& \sum_{n=0}^\infty
     (E-E_0)^n\beta_n(E_0,r)\ ,
\end{eqnarray}
where the unknown expansion coefficients $\alpha_n$ and $\beta_n$ are $(N\times N)$-matrices depending not only on variable $r$ but also on the choice of the
point $E_0$.\\

Eqs. (\ref{tAeq},\ref{tBeq}) involve matrices $\tilde{J}$ and $\tilde{Y}$
for which we can obtain the
expansions of the same kind but with known coefficients, $\gamma_n$ and $\eta_n$,
\begin{eqnarray}
\label{PSEJ}
     \tilde{J}(E,r) &=& \sum_{n=0}^\infty
     (E-E_0)^n\gamma_n(E_0,r)\ ,\\[3mm]
\label{PSEY}
     \tilde{Y}(E,r) &=& \sum_{n=0}^\infty
     (E-E_0)^n\eta_n(E_0,r)\ .
\end{eqnarray}
Simple recurrency relations for calculating the expansion coefficients
(matrices) $\gamma_n(E_0,r)$ and $\eta_n(E_0,r)$ are derived in the
Appendix \ref{appendix}.\\

Substituting the expansions (\ref{PSEA}, \ref{PSEB}, \ref{PSEJ}, \ref{PSEY})
into Eqs. (\ref{tAeq},\ref{tBeq}), and equating the coefficients of the same
powers of $(E-E_0)$, we obtain the following system of differential equations
for the unknown matrices $\alpha_n$ and $\beta_n$
\begin{eqnarray}
\label{aEq}
    \partial_r\alpha_n &=&
    -\sum_{i+j+k=n}\eta_iV(\gamma_j\alpha_k-\eta_j\beta_k)\ ,\\[3mm]
\label{bEq}
    \partial_r\beta_n &=&
    -\sum_{i+j+k=n}\gamma_iV(\gamma_j\alpha_k-\eta_j\beta_k)\ ,
    \qquad n=0,1,2,\dots
\end{eqnarray}
The boundary conditions (\ref{tABbcond}) are independent of $E$. As is easy
to see, this implies that
\begin{equation}
\label{anbncond}
    \alpha_n(E_0,0)=\delta_{n0}I\ ,\qquad
    \beta_n(E_0,0)=0\ .
\end{equation}
In other words, all these matrices vanish at the origin, except for the matrix
$\alpha_0$ which becomes a diagonal unit matrix at $r=0$.\\

Therefore for any finite radius $r$ and an arbitrary complex $E_0$, we have a
simple procedure for calculating the expansion coefficients. What we actually
need are the corresponding expansions of the matrices $\tilde{\cal A}(E,r)$ and
$\tilde{\cal B}(E,r)$ when $r\to\infty$, from which the Jost matrices
(\ref{JmAB}) can be constructed. In this context, taking the limit $r\to\infty$
is not a trivial procedure. Indeed, at large distances the Riccati-Bessel and
Riccati-Neumann functions behave as linear combinations of the exponential
functions (\ref{Hankel_ass}). There is nothing wrong in this behaviour if the
channel momenta are real. If however the energy is complex and the momenta have
non-zero imaginary parts, then the matrices $\tilde{J}(E,r)$ and
$\tilde{Y}(E,r)$ involve divergent exponential functions and thus are not
holomorphic when $r\to\infty$.\\

\begin{figure}[ht!]
\centerline{\epsfig{file=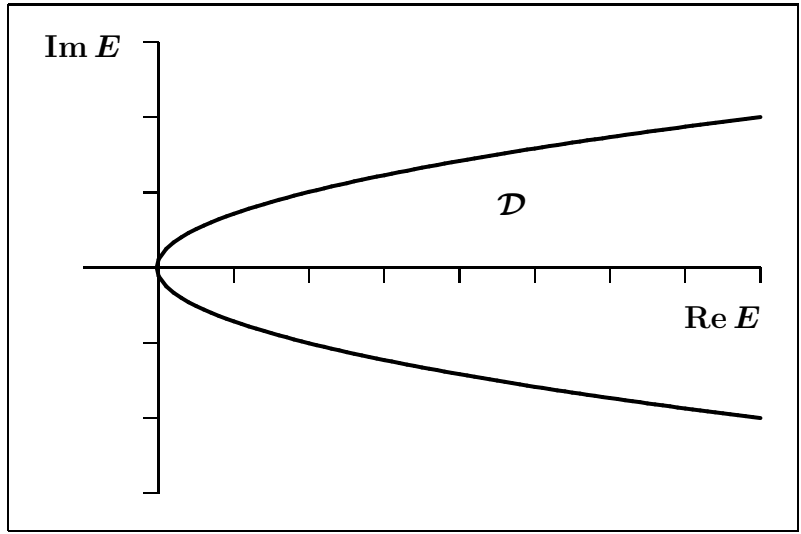}}
\caption{\sf
The domain $\cal D$ defined by Eq. (\ref{domainD}), for the two-channel model
described in Sec. \ref{sec.example}. The matrices $\tilde{\cal A}(E,\infty)$ and
$\tilde{\cal B}(E,\infty)$  for this model are holomorphic in the area enclosed
between the parabolic curves. The parabola crosses the real axis at
$E=-0.025$.
}
\label{fig.domainD}
\end{figure}

To some extent the situation can be saved by using exponentially decaying
potentials $V_{nn'}(r)\sim\exp(-\lambda_{nn'}r)$, which compensate the
divergence of $\tilde{J}(E,r)$ and $\tilde{Y}(E,r)$ within certain domain
$\cal{D}$ of the complex $E$-plane along its real axis. The borders of the
domain $\cal{D}$ are determined by the requirement that none of the coefficients
of Eqs. (\ref{tAeq},\ref{tBeq}) are divergent. The behaviour of these
coefficients is determined by the products
$exp(\pm ik_nr)\exp(-\lambda_{nn'}r)\exp(\pm ik_{n'}r)$. The worst case is when
we have the product of the most rapidly growing exponentials stemming from the
Riccati-Bessel and Riccati-Neumann functions, and the most slowly decaying
element of the potential matrix. Therefore
\begin{equation}
\label{domainD}
   {\cal D}=
   \left\{E:\max\limits_{n}\left|2{\rm Im\,}
   \sqrt{2\mu_n(E-E_n)/\hbar^2}\right|
   <\min\limits_{nn'}\lambda_{nn'}\right\}\ .
\end{equation}
The faster the potential matrix decays, the wider is the domain.
An example of such a domain is shown in Fig. \ref{fig.domainD} for the
two-channel model used in Sec. \ref{sec.example}.\\

Therefore when $r\to\infty$, we can use the Poincar\'e theorem only within $\cal
D$. This means that only within this domain the matrices $\tilde{\cal
A}(E,\infty)$ and $\tilde{\cal B}(E,\infty)$ are holomorphic functions of the
parameter $E$. A rigorous analysis (using a different method) of the analyticity
domain of the multi-channel $T$-matrix for the same class of potentials is given
by Motovilov in Ref. \cite{Motovilov}.\\

If the potential matrix $V(r)$ (or at least its long-range tail) is an analytic
function of complex variable $r$ and exponentially decays along any ray
$r=|r|e^{i\theta}$ within certain sector
$\theta_{\mathrm{min}}<\theta<\theta_{\mathrm{max}}$ of the complex $r$-plane,
then the domain of analyticity of the matrices $\tilde{\cal A}(E,\infty)$ and
$\tilde{\cal B}(E,\infty)$ can be extended by using the complex rotation
described in Sec. \ref{sec.rotation}.\\

The physically interesting domain of the $E$-plane where the expansion proposed
in the present paper, can be used in practical calculations, lies on the
positive real axis (scattering and reactions) and in the close vicinity  below
it (pronounced resonances). Therefore we can say that to all practical purposes
the expansion can be done near an arbitrary point $E_0$.\\

If we denote the asymptotic values of the matrices $\alpha_n$ and $\beta_n$ as
\begin{equation}
\label{abinfinity}
    \alpha_n(E_0,r)
    \ \mathop{\longrightarrow}\limits_{r\to\infty}
    \ a_n(E_0)\ ,
    \qquad\text{and}\qquad
    \beta_n(E_0,r)
    \ \mathop{\longrightarrow}\limits_{r\to\infty}
    \ b_n(E_0)\ ,
\end{equation}
then in vicinity of any chosen point $E_0$ on the Riemann surface, we can
obtain semi-analytic expressions for the Jost matrices in the form
\begin{eqnarray}
\label{fin_semi_exp}
    F^{\rm (in)}_{mn} &=&
                 \sum_{j=0}^M(E-E_0)^j\left[
                 \frac{k_n^{\ell_n+1}}{2k_m^{\ell_m+1}}(a_j)_{mn}-
                 \frac{ik_m^{\ell_m}k_n^{\ell_n+1}}{2}(b_j)_{mn}
                 \right]\ ,\\[3mm]
\label{fout_semi_exp}
    F^{\rm (out)}_{mn} &=&
                 \sum_{j=0}^M(E-E_0)^j\left[
                 \frac{k_n^{\ell_n+1}}{2k_m^{\ell_m+1}}(a_j)_{mn}+
                 \frac{ik_m^{\ell_m}k_n^{\ell_n+1}}{2}(b_j)_{mn}
                 \right]\ ,
\end{eqnarray}
where $M$ is the maximal power of our expansion. The non-analytic
quantities in Eqs. (\ref{fin_semi_exp}, \ref{fout_semi_exp}), i.e. the matrices
$a_j$ and $b_j$, are obtained by numerical integration of differential
equations (\ref{aEq}, \ref{bEq}) with the boundary conditions (\ref{anbncond})
from $r=0$ to certain large value $r=R$ (in some cases, this should be done
along the deformed contour shown in Fig. \ref{fig.pathray}). For a given energy
$E$, the choice of the sheet of the Riemann surface, where the Jost matrices are
considered, is done by appropriately choosing the signs in front of the square
roots (\ref{chmom_pm}) for calculating the channel momenta used in Eqs.
(\ref{fin_semi_exp}, \ref{fout_semi_exp}). The central point $E_0$ of the
expansion as well as the numerically obtained matrices $a_j$ and $b_j$ are the
same for all the layers of the Riemann surface.\\

In principle, the power-series expansions (\ref{fin_semi_exp},
\ref{fout_semi_exp}) includes infinite number of terms. In practice, however, we
may take into account just a few terms ($M<\infty$) and therefore have to solve
certain number of equations in the infinite system (\ref{aEq}, \ref{bEq}). It
should be emphasized that not all equations of this system are coupled to each
other. Indeed, due to the condition $i+j+k=n$ on their right-hand sides,
the equations for $\alpha_M$ and $\beta_M$ (for any $M\geqslant0$) are only
linked to the corresponding equations with $n<M$. For example, the first pair
of equations,
\begin{eqnarray}
\label{aEq0}
    \partial_r\alpha_0 &=&
    -\eta_0V(\gamma_0\alpha_0-\eta_0\beta_0)\ ,\\[3mm]
\label{bEq0}
    \partial_r\beta_0 &=&
    -\gamma_0V(\gamma_0\alpha_0-\eta_0\beta_0)\ ,
\end{eqnarray}
is self-contained and is not linked to any other equation of the system. The
second pair is linked only to the first one, and so on. This means that by
considering a finite number of equations of this system, we do not introduce a
truncation error.\\

In principle, it is possible to find some physical meaning (scattering length,
effective radius etc.) of certain combinations of the expansion coefficients
$(a_j)_{mn}$ and $(b_j)_{mn}$ in Eqs. (\ref{fin_semi_exp},\ref{fout_semi_exp}).
This would require the parametrization of the corresponding $S$-matrix
(\ref{S_matrix}) in terms of the channel phase-shifts. This is easily done for
a single channel problem (see Ref. \cite{my2009}). However for a multi-channel
system (even with minimal $N=2$) this results in rather complicated formulae
which do not add any clarity to the method. Moreover, the main advantage of the
proposed method consists in the fact that it can be used not only at low but at
any energies (even complex) where the notions of scattering length and effective
radius are meaningless anyway. This is why we think that within our method it is
not worthwhile to derive any expressions for something like multi-channel
scattering length and effective radius.

\section{Numerical example}
\label{sec.example}
\begin{figure}[ht!]
\centerline{\epsfig{file=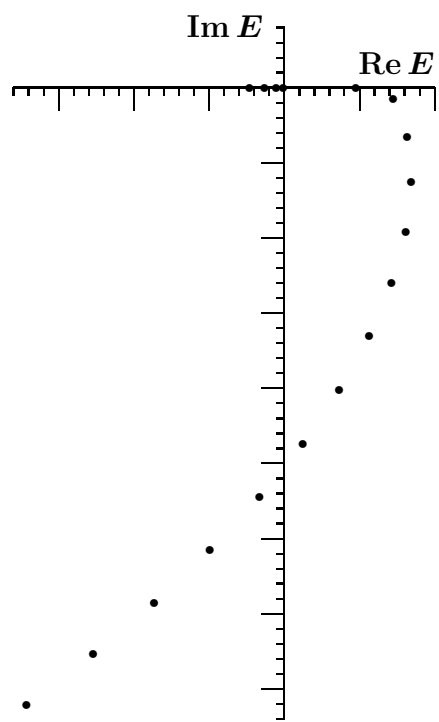}}
\caption{\sf
Spectral points generated by the potential (\ref{NTpotential}) and
given in  Table~\protect\ref{table.spectrum}.
}
\label{fig.spectrum}
\end{figure}

\begin{figure}[ht!]
\centerline{\epsfig{file=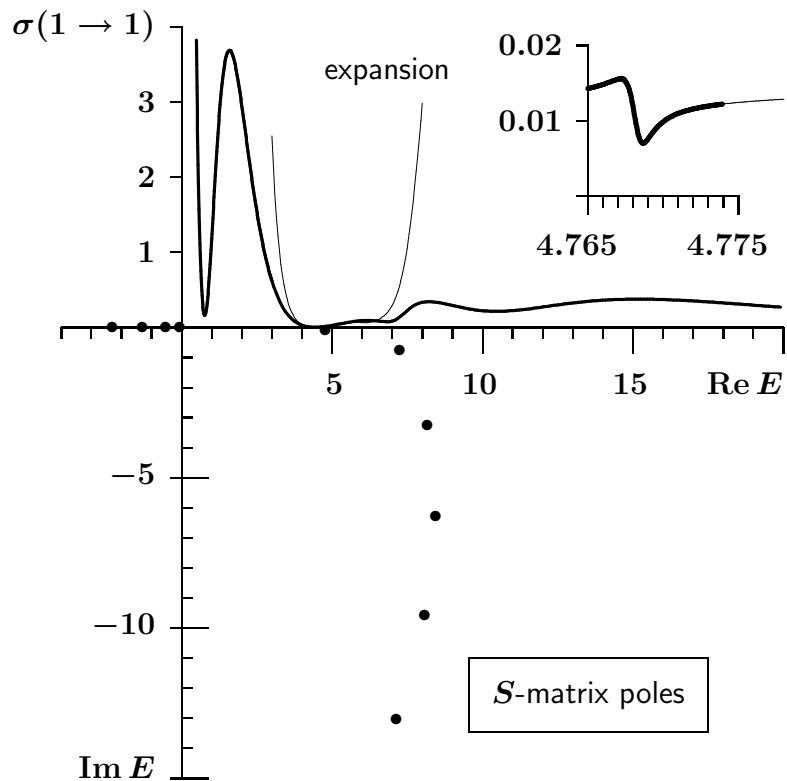}}
\caption{\sf
Energy dependence of the elastic scattering cross section in channel 1
for the potential (\ref{NTpotential}). Few of the $S$-matrix poles
(see Table~\protect\ref{table.spectrum} and
Fig.~\protect\ref{fig.spectrum}) are shown in the lower part of the
Figure. The thick curve represents the exact cross section, while the thin
curve shows the cross section obtained with the expansions
(\ref{fin_semi_exp}, \ref{fout_semi_exp}) where $E_0=5+i0$ and $M=5$.
}
\label{fig.s11}
\end{figure}

\begin{figure}[ht!]
\centerline{\epsfig{file=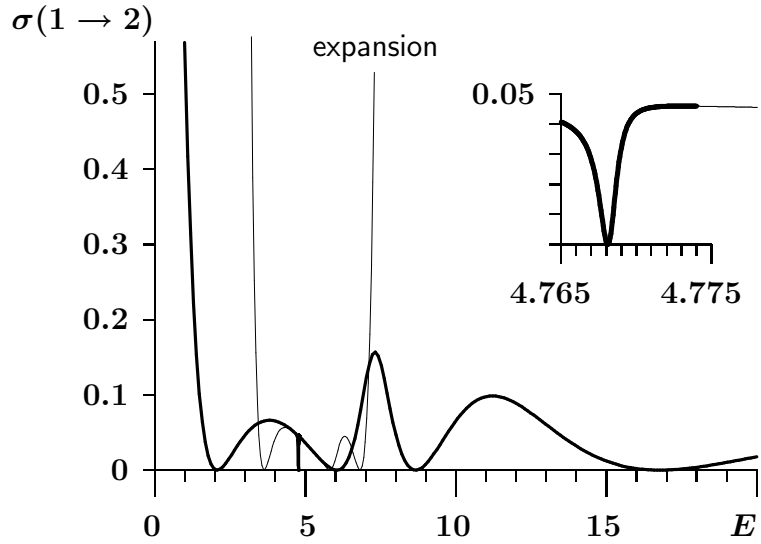}}
\caption{\sf
Cross section energy dependence of the inelastic transition
($1\to2$) for the potential (\ref{NTpotential}). The thick curve represents the
exact cross section, while the thin
curve shows the cross section obtained with the expansions
(\ref{fin_semi_exp}, \ref{fout_semi_exp}) where $E_0=5+i0$ and $M=5$.
}
\label{fig.s12}
\end{figure}

\begin{figure}[ht!]
\centerline{\epsfig{file=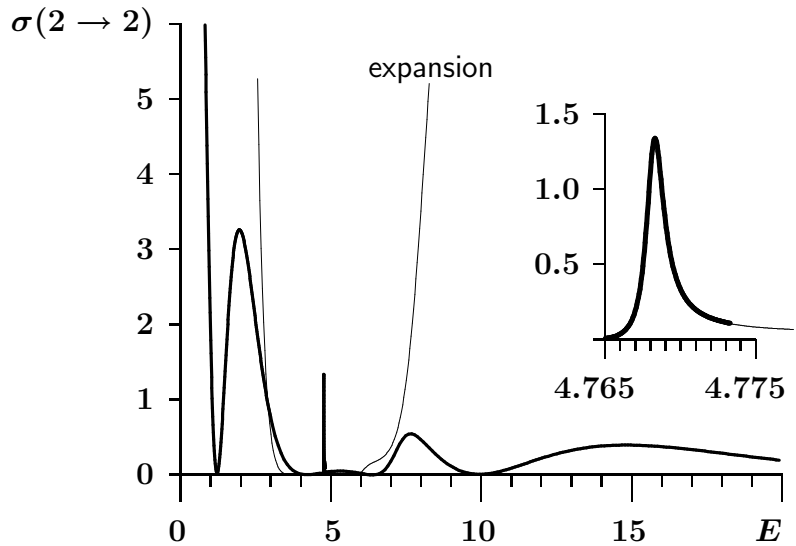}}
\caption{\sf
Energy dependence of the elastic scattering cross section in channel 2
for the potential (\ref{NTpotential}). The thick curve represents the exact
cross section, while the thin
curve shows the cross section obtained with the expansions
(\ref{fin_semi_exp}, \ref{fout_semi_exp}) where $E_0=5+i0$ and $M=5$.
}
\label{fig.s22}
\end{figure}

In order to demonstrate the efficiency and accuracy of the proposed method, we
do numerical calculations for a well-studied model. For this purpose, we use
the two-channel potential suggested by Noro and Taylor \cite{norotaylor}
\begin{equation}
\label{NTpotential}
    V(r)=\begin{pmatrix}
    -1.0 & -7.5\\
    -7.5 & 7.5\\
    \end{pmatrix}
    r^2e^{-r}\ ,
\end{equation}
which is given in arbitrary units such that $\mu_1=\mu_2=\hbar c=1$. The
threshold energies for the two channels of the Noro and Taylor model are $E_1=0$
and $E_2=0.1$, and the angular momentum is zero in both channels,
$\ell_1=\ell_2=0$. This potential has an attractive well in the first channel,
a repulsive barrier in the second, and rather strong coupling between the
channels. As a result, it generates a rich spectrum of bound and resonant
states (see Fig. \ref{fig.spectrum}) as well as a non-trivial energy dependence
of the channel and
transition cross sections (see Figs. \ref{fig.s11}, \ref{fig.s12}, and
\ref{fig.s22}) \cite{two_channel}.
This model is therefore a difficult testing ground for any new method designed
for describing multi-channel processes.\\

As a first test, we do the power series expansions of the Jost matrices
(\ref{fin_semi_exp}, \ref{fout_semi_exp}) around the point $E_0=5+i0$ on the
real axis and with six terms, i.e. with $M=5$. The central point of the
expansion was chosen to be not far from the first resonance, where the channel
cross sections have some non-trivial energy dependence. As is seen in  Figs.
\ref{fig.s11}, \ref{fig.s12}, and \ref{fig.s22}, the thin curves representing
the approximate cross sections, reproduce the corresponding exact cross sections
rather well within a wide energy interval.\\

\begin{figure}[ht!]
\centerline{\epsfig{file=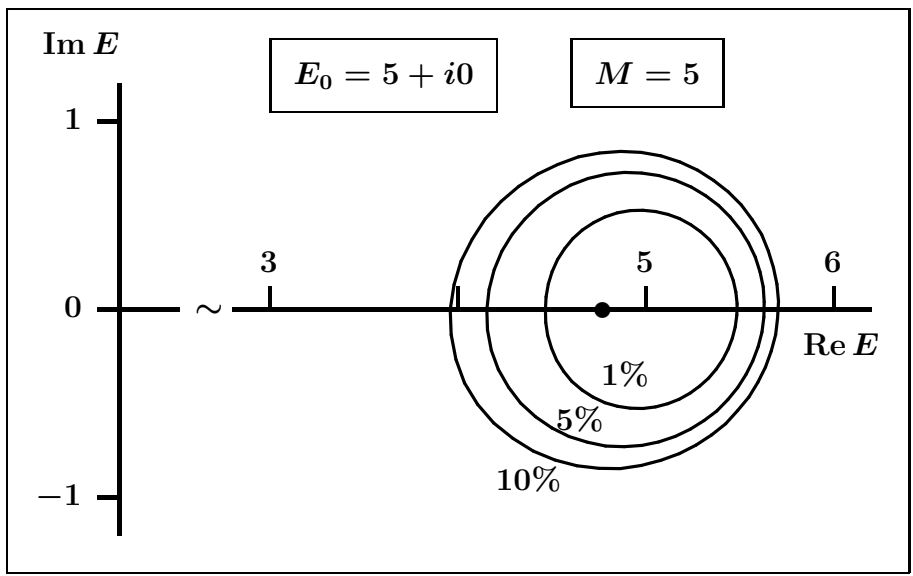}}
\caption{\sf
The domains within which the Jost matrix determinant
for the potential (\ref{NTpotential})
is reproduced, using the first five terms ($M=5$) of the expansion
(\ref{fin_semi_exp}), with the accuracy better than 1\%,
5\% and 10\%. The expansion was done around the point $E_0=5$ on the real axis.
}
\label{fig.domain5}
\end{figure}

Since the approximate Jost matrices  (\ref{fin_semi_exp}, \ref{fout_semi_exp})
coincide with the exact matrices on a segment of the real axis, they must be
valid also at the nearby points of the complex energy surface. Comparing the
exact and approximate values of $\det F^{\rm (in)}(E)$ at complex $E$ around
the point $E_0$ on the third sheet $(--)$ of the Riemann surface (see Fig.
\ref{fig.Riemann_double}), we
found the domains within which the relative accuracy is better than $1\%$,
$5\%$, and $10\%$. These domains are shown in Fig. \ref{fig.domain5}.\\

It is seen that the first resonance is within the domain of $1\%$ accuracy and
therefore must be reproduced by the approximate Jost matrix. And indeed, its
determinant has zero at $E=4.768178-i0.000686$ which is very close to the exact
value $E=4.768197-i0.000710$. With more terms in the expansion, the difference
becomes smaller, and with $M=10$ all the digits are the same.\\

\begin{figure}[ht!]
\centerline{\epsfig{file=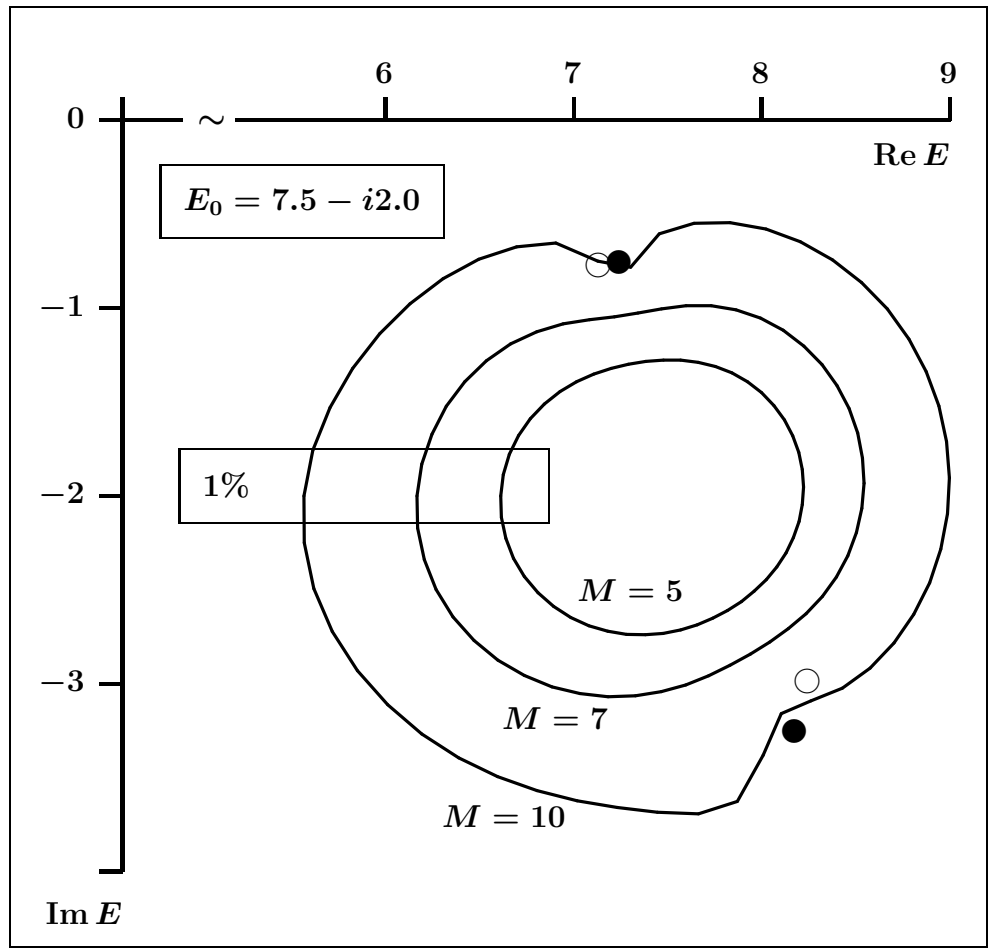}}
\caption{\sf
Growth of the 1\% accuracy domain with the increase of the number $M$
of terms in expansions (\ref{fin_semi_exp}), which were done around
$E_0=7.5-i2.0$ for the potential (\ref{NTpotential}). Filled circles indicate
the exact position of two resonances, while the open circles are their
approximate positions obtained with $M=5$.
}
\label{fig.domain_5_7_10}
\end{figure}

The other place where we tested the expansion, was the point $E_0=7.5-i2.0$ on
the third sheet $(--)$ of the Riemann surface. This point is almost in the
middle between the second and the third resonances. Fig. \ref{fig.domain_5_7_10}
shows how the domain of $1\%$ accuracy increases with increasing number of
terms ($M=5, 7, 10$) in the expansions (\ref{fin_semi_exp},
\ref{fout_semi_exp}). As is seen in Fig. \ref{fig.domain_5_7_10}, even with
$M=5$ both resonances are reproduced relatively well (the filled and open
circles represent the exact and approximate positions of the resonances). The
zeros of the exact Jost matrix determinat are at
$E=7.241200-i0.755956$ and $E=8.171217-i3.254166$ while the expansion with
$M=5$ gives $E=7.131204-i0.768670$ and $E= 8.241795-i2.982867$.
We deliberately chose the point $E_0$ far away from both resonances. If it is
close to any of them, the resonance can be found very accurately.

\section{Conclusion}
In the present paper, we show that each matrix element of multi-channel Jost
matrix can be written as a sum of two terms, and each term can be
factorized in such a way that it assumes the form of a product of certain
combination of the channel momenta $k_n$ times an analytic single-valued
function of the energy $E$. This means that all the branching points of the
Riemann energy-surface are given in the Jost matrix explicitly via the
channel-momentum factors. The remaining energy-dependent factors in all its
matrix elements are defined on single energy plane which does not have any
branching points anymore.\\

For these energy-dependent functions, we derive a system of first-order
differential equations. Then, using the fact that the functions are analytic,
we expand them in the power series and obtain a system of differential
equations that determine the expansion coefficients.
A systematic procedure developed in the present paper, allows us to accurately
calculate the power-series expansion of the Jost matrices practically at any
point on the Riemann surface of the energy. Actually, the expansion is done for
the single-valued functions of the energy, while the choice of the sheet of the
Riemann surface is done by appropriately choosing  the signs of the channel
momenta in the momentum-dependent factors.\\

The expansion suggested in the present paper, makes it possible to obtain a
semi-analytic expression for the Jost-matrix (and therefore for the S-matrix)
near an arbitrary point on the Riemann surface and thus to locate the spectral
points (bound and resonant states) as the S-matrix poles. Alternatively, the
expansion can be used to parametrize experimental data, where the unknown
expansion coefficients are the fitting parameters. Such a parametrization will
have the correct analytic properties. After fitting the data given at real energies, one can use the semi-analytic Jost matrix to search for resonances in the nearby domain of the Riemann surface. The efficiency and
accuracy of the suggested expansion is demonstrated by an example of a
two-channel model.
\appendix
\begin{center}
{\bf APPENDIX:}
\end{center}
\section{Expansion of the Riccati functions}
\label{appendix}
As is given by Eqs. (\ref{jseries}, \ref{yseries}), the Riccati-Bessel and
Riccati-Neumann functions can be factorized as
\begin{equation}
\label{Ajyseries}
    j_\ell(kr)=k^{\ell+1}\tilde{j}_\ell(E,r)\ ,
    \qquad
    y_\ell(kr)=k^{-\ell}\tilde{y}_\ell(E,r)\ ,
\end{equation}
where the tilded functions depend on $k^2$, i.e. on the energy.
These functions are holomorphic and thus can be expanded in Taylor series at an
arbitrary point $E=E_0$,
\begin{equation}
\label{tjyTaylor}
     \tilde{j}_\ell(E,r)=\sum_{n=0}^\infty
     (E-E_0)^ng_{\ell n}(E_0,r)\ ,
     \qquad
     \tilde{y}_\ell(E,r)=\sum_{n=0}^\infty
     (E-E_0)^nt_{\ell n}(E_0,r)\ ,
\end{equation}
where the coefficients are given by the derivatives
\begin{equation}
\label{jTaylorcoeff}
     g_{\ell n}(E_0,r)=\frac{1}{n!}\left[\frac{d^n}{dE^n}
     \tilde{j}_\ell(E,r)\right]_{E=E_0}=
     \frac{1}{n!}\left[\frac{d^n}{dE^n}
     \frac{j_\ell(kr)}{k^{\ell+1}}\right]_{E=E_0}\ ,
\end{equation}
\begin{equation}
\label{yTaylorcoeff}
     t_{\ell n}(E_0,r)=\frac{1}{n!}\left[\frac{d^n}{dE^n}
     \tilde{y}_\ell(E,r)\right]_{E=E_0}=
     \frac{1}{n!}\left[\frac{d^n}{dE^n}
     k^\ell y_\ell(kr)\right]_{E=E_0}\ .
\end{equation}
In order to find these derivatives, we use the following relations
\cite{abramowitz}
\begin{equation}
\label{jabramw}
     \frac{d}{dz}\left[\frac{j_\ell(z)}{z^{\ell+1}}\right]=
     -\frac{j_{\ell+1}(z)}{z^{\ell+1}}
     \qquad\text{and}\qquad
     \frac{d}{dz}\left[z^\ell y_\ell(z)\right]=z^\ell y_{\ell-1}(z)\ .
\end{equation}
After a simple but lengthy algebra, we finally obtain
\begin{equation}
\label{gcoeff}
     g_{\ell n}(E_0,r)=\frac{1}{n!}\left[\left(-\frac{\mu r}{\hbar^2}
     \right)^n\frac{j_{\ell+n}(kr)}{k^{\ell+n+1}}\right]_{E=E_0}\ ,
\end{equation}
\begin{equation}
\label{tcoeff}
     t_{\ell n}(E_0,r)=\frac{1}{n!}\left(\frac{\mu r}{\hbar^2}
     \right)^n\left[k^{\ell-n}y_{\ell-n}(kr)\right]_{E=E_0}\ .
\end{equation}
The matrices $\gamma_n$ and $\eta_n$ of Eqs. (\ref{PSEJ}, \ref{PSEY}) are
diagonal with each row having the functions (\ref{gcoeff},
\ref{tcoeff}) with $\mu$ and $k$ for the corresponding channel. These functions
should be the same for all sheets of the Riemann surface, i.e. for any choice of
the signs of channel momenta. This is so indeed since
$j_\ell(-z)=(-1)^{\ell+1}j_\ell(z)$ and $y_\ell(-z)=(-1)^\ell y_\ell(z)$.


%
%
\begin{table}
\begin{center}
\begin{tabular}{|r|r|r|r|}
\hline
$E_r$ & $\Gamma$ & $\Gamma_1$ & $\Gamma_2$\\
\hline
 -2.314391 & 0 & 0 & 0\\
\hline
 -1.310208 & 0 & 0 & 0\\
\hline
 -0.537428 & 0 & 0 & 0\\
\hline
 -0.065258 & 0 & 0 & 0\\
\hline
4.768197 & 0.001420 & 0.000051 & 0.001369\\
\hline
7.241200 & 1.511912 & 0.363508 & 1.148404\\
\hline
8.171217 & 6.508332 & 1.596520 & 4.911812\\
\hline
8.440526 & 12.562984 & 3.186169 & 9.376816\\
\hline
8.072643 & 19.145630 & 4.977663 & 14.167967\\
\hline
7.123813 & 26.025337 & 6.874350 & 19.150988\\
\hline
5.641023 & 33.070140 & 8.816746 & 24.253394\\
\hline
3.662702 & 40.194674 & 10.768894 & 29.425779\\
\hline
1.220763 & 47.339350 & 12.709379 & 34.629971\\
\hline
-1.657821 & 54.460303 & 14.624797 & 39.835506\\
\hline
-4.949904 & 61.523937 & 16.507476 & 45.016461\\
\hline
-8.635366 & 68.503722 & 18.352084 & 50.151638\\
\hline
-12.696283 & 75.378773 & 20.155213  & 55.223560\\
\hline
-17.117760 & 82.129712 & 21.915313 & 60.214399\\
\hline
\end{tabular}
\end{center}
\caption{
Spectral points $E=E_r-i\Gamma/2$ generated by the potential (\ref{NTpotential})
and
shown in  Fig.~\protect\ref{fig.spectrum}.
}
\label{table.spectrum}
\end{table}

%
%

\end{document}